\documentclass[12pt]{article}
 
\usepackage{amsmath}
\usepackage{amsthm}
\usepackage{amssymb}
\usepackage{amsfonts}

\usepackage[latin1]{inputenc}

\usepackage{graphicx}

\def\be{\begin{equation}}
\def\ee{\end{equation}}
\def\bea{\begin{eqnarray}}
\def\eea{\end{eqnarray}}

\newcommand{\bq}{\mathbf{q}}
\newcommand{\bp}{\mathbf{p}}

\newcommand{\te}{\theta}

\newcommand{\hbq}{\hat{\mathbf{q}}}
\newcommand{\hbp}{\hat{\mathbf{p}}}
\newcommand{\hH}{\hat{\cal{H}}}
\newcommand{\hC}{\hat{C}}
\newcommand{\hI}{\hat{I}}

\newcommand{\hq}{\hat{q}}
\newcommand{\hp}{\hat{p}}

\def\RR{\mathbb{R}}

\def\dd{{\rm d}}
\newcommand\Om\Omega
\newcommand{\bL}{\mathbf{L}}

\newcommand{\hbL}{\hat{\mathbf{L}}}

\newcommand\minus\backslash

\newcommand{\om}{\omega}

\def\la{{\lambda}}
\def\ma{{\mu}}
\def\rmi{{\rm i}}
\def\rme{{\rm e}}

\def\1{\'{\i}}

\def\k{{\kappa}}

\def\>#1{{\mathbf#1}}

\def\omm{\Omega}

\def\bb{\beta}

\def\arc{{\rm arc}\!}

\begin{document}

 \
 \smallskip

  {\large{\bf{A new exactly solvable quantum model 
in N dimensions }}}

\bigskip

\begin{center}
{\sc \'Angel Ballesteros$^a$,   Alberto Enciso$^b$,  Francisco J. Herranz$^a$,\\[4pt] Orlando Ragnisco$^c$ and Danilo Riglioni$^c$}

\bigskip

 \small

{$^a$ Departamento de F\1sica,  Universidad de Burgos,
09001 Burgos, Spain\\ ~~E-mail: angelb@ubu.es \quad fjherranz@ubu.es\\[10pt]
}
$^b$ Departamento de F\1sica Te\'orica II,   Universidad Complutense,   28040 Madrid,
Spain ~~E-mail: aenciso@fis.ucm.es\\[10pt]
$^c$ Dipartimento di Fisica,   Università di Roma Tre and Istituto Nazionale di
Fisica Nucleare sezione di Roma Tre,  Via Vasca Navale 84,  00146 Roma, Italy  \\
~~E-mail: ragnisco@fis.uniroma3.it  \quad riglioni@fis.uniroma3.it    \\[10pt]
\end{center}

\medskip
\medskip

\begin{abstract}
\noindent The $N$-dimensional position-dependent mass Hamiltonian
$$
\hat H = \frac{-\hbar^2}{2(1+\lambda \mathbf{q}^2)} \nabla^2 +\frac{\omega^2 \mathbf{q}^2}{2(1+\lambda \mathbf{q}^2)}
$$
is shown to be exactly solvable for any real positive value of the parameter $\lambda$. Algebraically, this Hamiltonian can be thought of as a new maximally superintegrable $\la$-deformation of the $N$-dimensional isotropic oscillator and, from a geometric viewpoint, this system is just the intrinsic oscillator potential on an $N$-dimensional hyperbolic space with nonconstant curvature.
The spectrum of this model is shown to be hydrogenlike, and their eigenvalues and eigenfunctions are explicitly obtained by deforming appropriately the symmetry properties of the $N$-dimensional harmonic oscillator. A further generalization of this construction giving rise to new exactly solvable models is envisaged.

 \end{abstract}

\medskip
\medskip



\newpage

%%%%%%%%%%%%%%%%%%%%%%%%%%%%%%%%%%%%%%%%%%%%%%%%%%%
 
\section{Introduction}

 The $N$-dimensional ($N$D) classical Hamiltonian given by
  \be 
  {\cal H}(\bq,\bp)={\cal T}(\bq,\bp)+{\cal U}(\bq)=\frac{\bp^2}{2(1+\la \bq^2)} +
   \frac{ \om^2 \bq^2}{2(1+\la \bq^2)} ,
   \label{ac}
\ee 
  with  real parameters $\la>0$ and $\om\geq 0$ and where $\bq,\bp\in\RR^N$ are conjugate coordinates and momenta,  was proven in~\cite{PhysD} to be maximally superintegrable. This means that ${\cal H}$ is endowed with the maximum possible number of  $(2N-1)$ functionally independent constants of motion. 
Explicitly, the integrals of the motion for ${\cal H}$ are the ones that encode the radial symmetry of the system, namely,
\be
  C^{(m)}=\!\! \sum_{1\leq i<j\leq m} \!\!\!\! (q_ip_j-q_jp_i)^2 , \quad 
 C_{(m)}=\!\!\! \sum_{N-m<i<j\leq N}\!\!\!\!\!\!  (q_ip_j-q_jp_i)^2 , \quad m=2,\dots,N;\label{af}
 \ee
 together with the functions
 \be
 I_i=p_i^2-\bigl(2\la  {\cal H}(\bq,\bp)-\om^2\bigr) q_i^2 ,\qquad i=1,\dots,N.
\label{ag}
\ee
 Moreover, each of the three  sets $\{{\cal H},C^{(m)}\}$,  
$\{{\cal H},C_{(m)}\}$ ($m=2,\dots,N$) and   $\{I_i\}$ ($i=1,\dots,N$) is  formed by $N$ functionally independent functions  in involution, and the set $\{ {\cal H},C^{(m)}, C_{(m)},  I_i \}$ for $m=2,\dots,N$ with a fixed index $i$ provides the set of $(2N-1)$ functionally independent functions. Alternatively to the approach performed in~\cite{PhysD}, both sets of integrals (\ref{af}) and (\ref{ag}) can also be obtained through a St\"ackel transform (or coupling constant metamorphosis)~\cite{Hietarinta,Stackel2,Stackel4,Sergyeyev, Kalnins1} from the free Euclidean motion which has been achieved in~\cite{IJTP}.

Maximally superintegrable Hamiltonians in $N$ dimensions are quite scarce~\cite{annals} even on the Euclidean space, and the two representative examples of this class of systems with periodic bounded trajectories are the Kepler system and the isotropic harmonic oscillator. In fact, $  {\cal H}$  (\ref{ac}) can be interpreted as a genuine (maximally superintegrable) $\la$-deformation of the $N$D Euclidean isotropic oscillator with frequency $\om$, since the limit  $\la\to 0$ of (\ref{ac}) yields 
$$
 {\cal H}_0=\frac 12 \bp^2+\frac 12 \om^2\bq^2  .
$$
Moreover, an important property of ${\cal H}$ is that it can be written as
$$
{\cal H}=\frac 12 \sum_{i=1}^N I_i ,
$$
and, again, the limit $\la\to 0$ of the integrals $I_i$ transforms this equation into the separability property for the $N$D harmonic oscillator.

 From a geometric perspective, $ {\cal T}$ can be interpreted as the kinetic energy  defining the geodesic motion of a particle with unit mass  on a conformally flat space which is   the complete Riemannian manifold $ \RR^N$, with metric 
\be
\dd s^2= (1+\la \bq^2)\dd \bq^2 ,
\label{metric1}
\ee
and nonconstant scalar curvature  given by
$$
 R=-\la\,\frac{(N-1)\bigl( N(2+3\la \bq^2)-6\la \bq^2\bigr)}{(1+\la \bq^2)^3} .
$$
 The scalar curvature $R(r)\equiv R(|\bq|)$  is always a  {\em negative} increasing function such that 
 $\lim_{r\to \infty}R=0$ and  it has a minimum at the origin $$R(0)=-2\la N(N-1),$$ which is exactly the scalar curvature of the $N$D {\em hyperbolic space} with negative constant sectional curvature equal to $-2\la$.
In fact, such a curved space is the $N$D spherically symmetric generalization of the Darboux surface of type III~\cite{Ko72,KKMW03}, which was constructed in~\cite{annals, PLB}. On the other hand,   the central potential $ {\cal U}$ was proven in~\cite{PhysD,annals} to be an ``intrinsic" oscillator potential on that Darboux space (we remark that in \cite{PhysD} we considered the Hamiltonian $H$ with 
  ${\cal H}=  \k H/ 2$ and  $\la=1/\k$).

For $N=3$,  the potential $\cal U$  appears in the context of the so called Bertrand spacetimes~\cite{Perlick, Bertrand}. These spaces are  certain $(3+1)$D static and spherically symmetric Lorentzian spacetimes for which their bounded geodesic motions are all   periodic and possess stable circular orbits. Therefore these spaces provide the natural arena for the generalization of  the classical Bertrand's theorem~\cite{Bertrand2} to relativistic spaces of nonconstant curvature, and the maximal superintegrability of the associated 3D Bertrand Hamiltonians has been recently proven in~\cite{commun}.
 Again for $N=3$, $\cal H$  arises as a particular   case of the generalizations of  the MIC--Kepler and   Taub-NUT systems constructed in~\cite{IK95, uwano} (in particular, $\cal H$ can be recovered by setting $\nu=1/2$, $a=1$ and $b=\la$).
     
Alternatively, $\cal H$  can be interpreted as a position-dependent mass system in which the conformal factor of the metric (\ref{metric1})  is identified with the variable mass function
$$
 m(\bq)=1+\la \bq^2.
\label{aadd}
$$
The construction and analysis of the wide range of applications of several position-dependent mass Schr\"odinger Hamiltonians can be found in, for instance,  \cite{Roos}--\cite{MR}  (see references therein). In particular, one-dimensional  models containing quadratic mass functions are considered in~\cite{Koc, Schd} for certain semiconductor heterostructures.

In this letter we   solve the quantization problem for $\cal H$ when $\la>0$. In the next section we   introduce the radial effective potential associated to this system. In section 3 we   perform the quantization by imposing the existence of the quantum analog of the full set of integrals of the motion. The explicit solution of the spectral problem is presented in section 4. Furthermore, a generalization of this approach in order to obtain new exactly solvable quantum models with position-dependent mass is sketched in the last section.

 %%%%%%%%%%%%%%%%%%%%%%%%%%%%%%%%%%%%%%%%%%%%%%%%%%%
 
 \section{The classical effective potential}

 The Hamiltonian $\cal H$  (\ref{ac}) can also be expressed in terms of hyperspherical coordinates $r,\te_j$, and canonical   momenta $p_r,p_{\te_j}$,   $(j=1,\dots,N-1)$.
 The $N$ hyperspherical coordinates are given by the radial one $r=|\bq|\in \RR^+$  and   $N-1$   angles $\theta_j\in[0,2\pi)$. In terms of them we have that
$$
q_j=r \cos\te_{j}     \prod_{k=1}^{j-1}\sin\te_k ,\quad 1\leq j<N,\qquad 
q_N =r \prod_{k=1}^{N-1}\sin\te_k ,
$$
where hereafter any product $\prod_{l}^m$ such that $l>m$ is assumed to be equal to 1.
The metric (\ref{metric1}) now reads
\begin{equation}
\dd s^2= (1+\la r^2)(\dd r^2+r^2\dd\Om^2),
\label{bb}
\end{equation}
where   $\dd\Om^2$  is the   metric on the unit $(N-1)$D sphere
$$
\dd\Om^2=\sum_{j=1}^{N-1}\dd\te_j^2\prod_{k=1}^{j-1}\sin^2\te_k .
$$

Thus the Hamiltonian (\ref{ac}) becomes
  \be
 {\cal H}(r,p_r)={\cal T}(r,p_r)+{\cal U}(r) = 
 \frac{p_r^2+r^{-2}\bL^2 }{2(1+\la r^2)} + \  \frac{ \om^2 r^2}{2(1+\la r^2)} ,
 \label{bg}
 \ee
 where  the total   angular momentum  is given by
$$
\bL^2=\sum_{j=1}^{N-1}p_{\te_j}^2\prod_{k=1}^{j-1}\frac{1}{\sin^{2}\te_k}.
$$
In these coordinates the integrals of motion $C_{(m)}$  (\ref{af})   adopt a compact form (the remaining $C^{(m)}$  and $I_i$  have more cumbersome expressions):
$$
C_{(m)}=\sum_{j=N-m+1}^{N-1}p_{\te_j}^2\prod_{k=N-m+1}^{j-1}\frac 1{\sin^{2}\te_k  },\quad m=2,\dots,N;
$$
so that  $C_{(N)} =\bL^2$, which is just the second-order Casimir of the $\mathfrak{so}(N)$-symmetry algebra of any central potential.

\begin{figure}
\includegraphics[height=7cm]{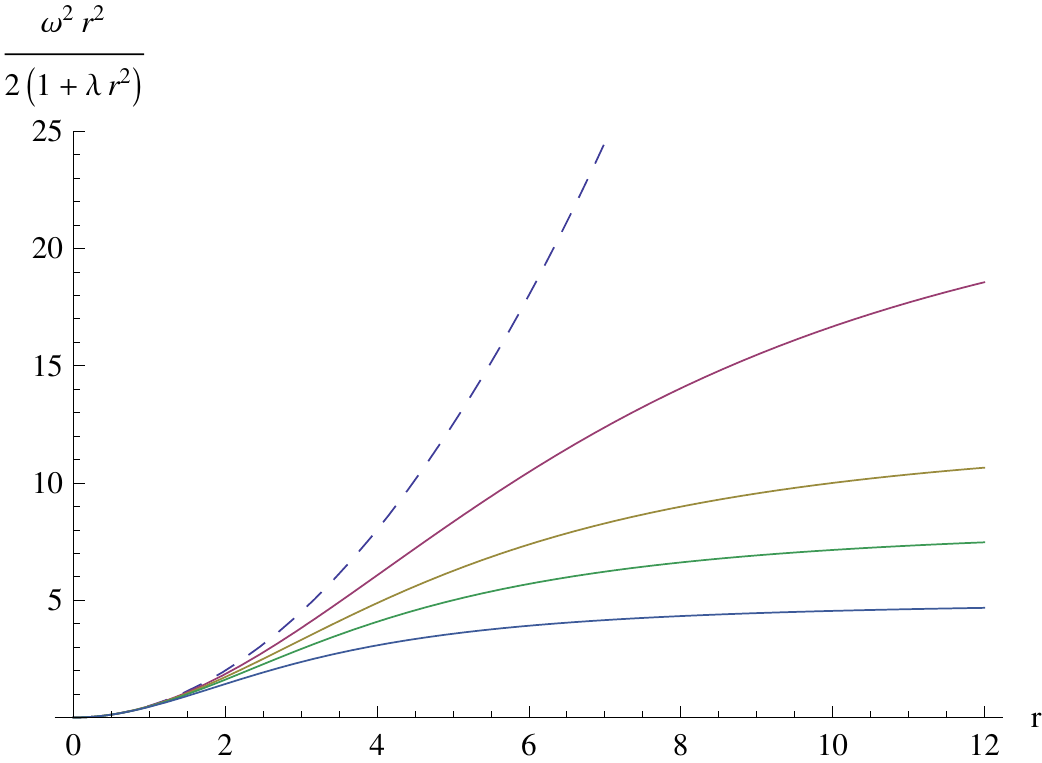}
\caption{The {nonlinear oscillator potential}   for $\la=\{0,\, 0.02,\, 0.04,\, 0.06,\, 0.1\}$ with $\omega=1$ is plotted. The upper   dashed  line corresponds to the isotropic harmonic oscillator  with $\la=0$, and the limit $r\to \infty$ gives   $\{ +\infty, \,25,\,12.5,\, 8.33,\, 5\}$, respectively.
 \label{fig1}}
\end{figure}

The nonlinear radial potential ${\cal U}(r)$  is always a  {\em positive} growing function for  $\la>0$ (see figure \ref{fig1})  and such that
\be
{\cal U}(0)=0 ,\qquad \lim_{r\to \infty}{\cal U}(r)=\frac{\om^2}{2\la} .
\label{bx}
\ee
However, we have to consider the {\em variable mass contribution} in the whole dynamics and this  can be better understood by introducing the radial effective  potential. If we apply the  canonical transformation defined by
$$
P(r,p_r)=\frac{p_r}{\sqrt{1+\la r^2}} ,\qquad Q(r)=\frac 12 r\sqrt{1+\la r^2}+\frac{\arc\sinh(\sqrt{\la}r)}{2\sqrt{\la}},
$$
on the 1D radial Hamiltonian (\ref{bg}), we get
  \be
 {\cal H}(Q,P)= \frac 12 P^2+ {\cal U}_{\rm eff}(Q ),\qquad
  {\cal U}_{\rm eff}(Q(r)  )= \frac{ c_N }{2(1+\la r^2)r^2} + \  \frac{ \om^2 r^2}{2(1+\la r^2)},
 \label{bbgg}
 \ee
where $c_N\ge 0$ is the value of the integral of motion corresponding to the total angular momentum $C_{(N)}\equiv \bL^2 $.
Hence the radial motion for the classical system can be described as a particle moving on  a 1D flat space under an effective potential $ {\cal U}_{\rm eff}(Q (r))$.
The analysis of ${\cal U}_{\rm eff}$ shows that,  for any value of $\la$, the energy of the system is always positive and it does have a minimum located at $r_{\rm min}$,
\be
r^2_{\rm min}=\frac{\la c_N+\sqrt{\la^2 c_N^2+\om^2 c_N}}{\om^2},\qquad    {\cal U}_{\rm eff}(Q (r_{\min}))=-\la c_N+\sqrt{\la^2 c_N^2+\om^2 c_N} ,
\label{ya}
\ee
 which for the harmonic oscillator $(\la=0)$ gives
$ r^2_{\rm min}=\frac{ \sqrt{  c_N}}{\om},\,{\cal U}_{\rm eff}(Q (r_{\min}))= \om \sqrt{  c_N}$.

\begin{figure}
\includegraphics[height=7cm,width=10cm]{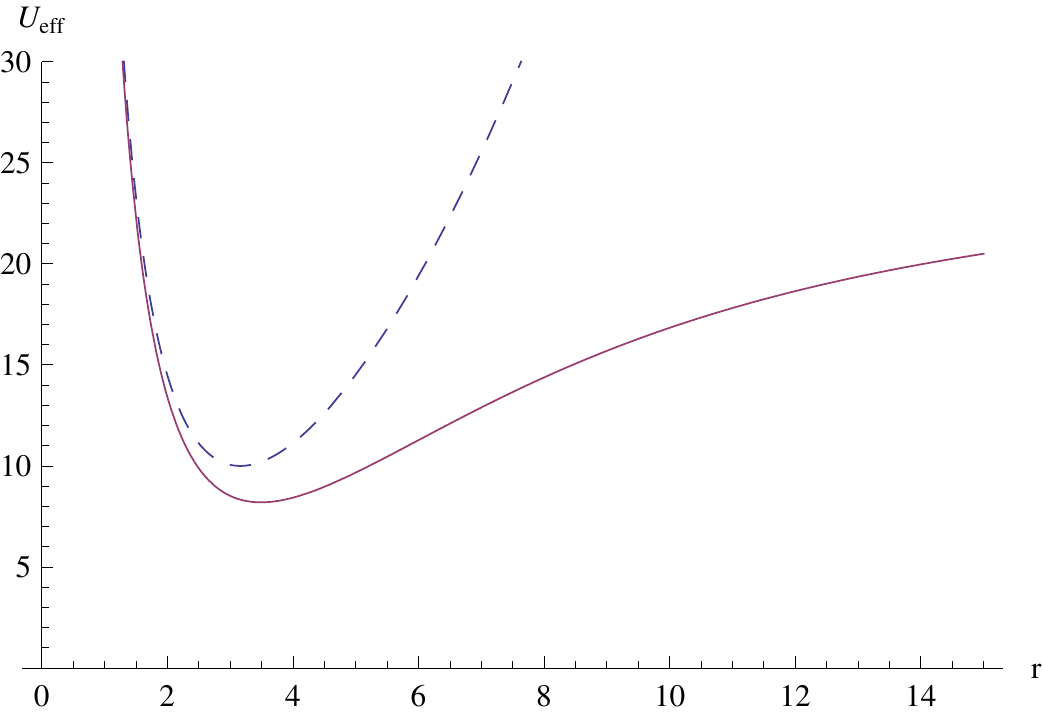}
\caption{Plot for the  {effective  nonlinear   oscillator potential} (\ref{ya})  for    $\la=0.02$, $c_N=100$ and $\om=1$. The minimum of the potential is located at $r_{\rm min}=3.49$ with ${\cal U}_{\rm eff}(r_{\rm min})=8.2$ and ${\cal U}_{\rm eff}(\infty)=25$.  The dashed  line   corresponds to the effective potential of the    harmonic oscillator  with $\la=0$ with minimum ${\cal U}_{\rm eff}(r_{\rm min})=10$ at  $r_{\rm min}=3.16$. 
 \label{fig2}}
\end{figure}

Now, if $\la>0$ and $c_N\neq 0$, then both $r,Q\in [0,\infty)$ and the effective potential has  two
representative limits:
\be
\lim_{r\to 0} {\cal U}_{\rm eff}(Q (r))=+\infty,\qquad \lim_{r\to \infty} {\cal U}_{\rm eff}(Q (r))=\frac{\om^2}{2\la} ,
\label{bz}
\ee
and the latter is just (\ref{bx}).  Thus, this effective potential turns out to be hydrogen-like. The values for the minimum of the effective potential, $r_{\rm min}$ and $  {\cal U}_{\rm eff}(Q (r_{\min}))$ (\ref{ya}) are, respectively, greater and smaller than those corresponding to the harmonic oscillator ($\la=0$) case. This effective potential is shown in figure \ref{fig2}. 

%%%%%%%%%%%%%%%%%%%%%%%%%%%%%%%%%%%%%%%%%%%%%

 \section{A maximally superintegrable quantization}
 
In order to obtain the quantum analogue of the kinetic energy term  (\ref{ac}) we have to deal with the unavoidable ordering problems in the canonical quantization process that come from the nonzero curvature of the underlying space (see, e.g.~\cite{Kalnins1}) or, equivalently, from its  alternative interpretation  as  a position-dependent mass Hamiltonian~\cite{Roos}--\cite{MR}.
 
A detailed analysis of  the different possible quantization prescriptions and a proof of their equivalence through gauge transformations will be presented in~\cite{BHERR}. One of these prescriptions consists in  the so called ``Schr\"odinger  quantization", which entails  that  the quantum Hamiltonian $\hH$ keeps the maximal superintegrability property and is  therefore endowed with $2N-1$ algebraically independent operators that commute with $\hH$. This prescription allows us to use the full symmetry machinery coming from the harmonic oscillator, and gives rise to the following main result.

 \noindent
{\bf Theorem 1.} {\em 
    Let   $\hH$ be the quantum Hamiltonian given by
 \be
 {\hH}= 
 \frac{1}{2(1+\la \hbq^2)}\, \hbp^2+ \  \frac{ \om^2 \hbq^2}{2(1+\la \hbq^2)}  ,
 \label{ca}
 \ee
 such that $[\hq_i,\hp_j]=\rmi \hbar \delta_{ij}$.
 For any real value of $\la$    it is verified that\\
\smallskip
(i) $\hH$ commutes with the following observables,
\be
  \hC^{(m)}=\!\! \sum_{1\leq i<j\leq m} \!\!\!\! ( \hq_i \hp_j- \hq_j \hp_i)^2  , \quad 
  \hC_{(m)}=\!\!\! \sum_{N-m<i<j\leq N}\!\!\!\!\!\!  ( \hq_i \hp_j- \hq_j \hp_i)^2  , \quad m=2,\dots,N;\label{cb}
 \ee
 \be
  \hI_i= \hp_i^2- 2\la  \hq_i^2 { \hH}( \hbq, \hbp)+ \om^2 \hq_i^2 ,\qquad i=1,\dots,N;
\label{cc}
\ee
where $\hC^{(N)}=\hC_{(N)}$ and ${\hH}=\frac 12 \sum_{i=1}^N \hI_i$.  
\smallskip\\
(ii) Each of the three  sets $\{{\hH},\hC^{(m)}\}$,  
$\{{\hH},\hC_{(m)}\}$ ($m=2,\dots,N$) and   $\{\hI_i\}$ ($i=1,\dots,N$) is  formed by $N$ algebraically independent  commuting observables.
\smallskip\\
(iii) The set $\{ { \hH},\hC^{(m)}, \hC_{(m)},  \hI_i \}$ for $m=2,\dots,N$ with a fixed index $i$    is  formed by $2N-1$ algebraically independent observables. \\
(iv) $\hH$ is formally self-adjoint on the Hilbert space $L^2(\RR^N,(1+\la\bq^2){\dd}\bq  )$, endowed with the scalar product
\[
\langle \Psi | \Phi \rangle_\la = \int_{\RR^N} \overline{{\Psi}(\bq)} \Phi(\bq)(1+\la\bq^2){\dd}\bq .\]
}

The proof of this result can be obtained through direct computation. Further details can be found in~\cite{BHERR}. 
Next, under 
the usual differential representation given by
$$
 \hq_i=q_i,\qquad \hp_i=-\rmi  \hbar \partial_i=-\rmi  \hbar \frac{\partial}{\partial q_i},\qquad \nabla=(\partial_1,\dots,\partial_N),$$ 
  the Hamiltonian (\ref{ca}) leads to the following Schr\"odinger equation
\begin{equation}
 \left( \frac{-\hbar^2}{2(1+\lambda \mathbf{q}^2)} \nabla^2 +\frac{\omega^2 \mathbf{q}^2}{2(1+\lambda \mathbf{q}^2)}\right)\Psi(\bq)=E\Psi(\bq) ,
 \label{kb}
\end{equation}  
 and the change to hyperspherical coordinates  yields    the equation
 \be
\frac{1 }{2(1+\la r^2)}\left(  {-\hbar^2}\partial_r^2 -\frac{\hbar^2(N-1)}{r}\,\partial_r + \frac{\hat{\mathbf{L}}^2}{r^2} + \omega^2 r^2  \right)\Psi(r,\te) = E\Psi(r,{\te}) ,
 \label{lh}
 \ee
where $\te=(\te_1,\dots,\te_{N-1})$ and   $\hbL^2\equiv \hC_{(N)}$ is the total  quantum  angular momentum.

 %%%%%%%%%%%%%%%%%%%%%%%%%%%%%%%%%%%%%%%%%%%%%%%%%%%
 
 \section{Spectrum and eigenfunctions}
 
From the effective potential (\ref{bbgg}) one should expect that the quantum Hamiltonian  (\ref{ca}) would have both a discrete and a continuous spectrum, and this is indeed  the case.

By taking into account that   ${\hH}$  can be defined in terms of the first integrals $\hat{I}_i$ through  ${\hH}=\frac 12 \sum_{i=1}^N \hI_i$,  we find that the discrete spectrum for the Schr\"odinger equation, $\hH \Psi(\bq)=E\Psi(\bq)$ (\ref{kb})  can  be fully determined by following exactly the same steps as in the {\em flat} isotropic  harmonic oscillator. Let us consider a factorized  wave function  together with the eigenvalue equations for the operators $\hI_i$:
$$
\Psi(\bq)=\prod_{i=1}^N \psi_i(q_i), \qquad \frac 12 \hI_i\Psi =\ma_i\Psi,\quad i=1,\dots,N.
$$
Since $\hI_i=\hI_i( \hq_i,\hp_i,\hH) $, we get
\be
\frac 12 \hI_i \psi_i (q_i)= \frac{1}{2}\left(-\hbar^2\partial_i^2 + (\omega^2-2\lambda E)q_i^2\right)\psi(q_i)= \ma_i \psi(q_i) . 
\label{rb}
\ee
Therefore if we define the frequency  
\be
\omm(E)=\sqrt{\om^2-2\la E}\quad {\rm whenever} \quad \om^2>2\la E,
\label{rc}
\ee
then (\ref{rb}) can be expressed, in a formal manner,  as the Schr\"odinger equation of the  one-particle harmonic oscillator 
\be
 \frac{1}{2}\left(-\hbar^2\partial_i^2 + \omm^2q_i^2\right)\psi(q_i)= \ma_i \psi(q_i) .
\ee
Since necessarily 
$ \psi(q_i)\in L^2(\RR, \dd q_i)$,  the eigenvalue $\ma_i$ and the wave function $\psi(q_i)$ turn out to be
\bea
&& \!\!\!\!\!\!\!\! \!\!\!\!\!\!\!\! \ma_i\equiv \ma_i(E ,n_i)=\hbar\omm\left(n_i+\frac 12\right) ,\qquad n_i=0,1,2\dots \nonumber\\
&&\!\!\!\!\!\!\!\! \!\!\!\!\!\!\!\! \psi_i(q_i)\equiv \psi_{n_i}(E,q_i)=A_{n_i} \left( \frac{\bb^2}{\pi}\right)^{1/4}\!\!\!\!\exp\{-\bb^2 q_i^2/2\} H_{n_i}(\bb q_i) ,\quad \bb=\sqrt{\frac{\omm}{\hbar}},
\label{rd}
\eea
where $H_{n_i}(\bb q_i)$ is the $n_i$-th Hermite polynomial and $A_{n_i}$ is a normalization constant.
 Now from
$$
\hH\Psi=\frac 12\left( \sum_{i=1}^N\hI _i\right) \Psi =  \sum_{i=1}^N \ma_i\Psi \equiv  E\Psi ,
$$
we obtain that
\be
E=\hbar\omm\left(n+\frac N2\right)=\hbar\sqrt{\om^2-2\la E}\left(n+\frac N2\right),\quad n=\sum_{i=1}^N n_i,\quad n=0,1,2\dots
\label{rree}
\ee
 which gives   an explicit expression for the energies 
\begin{equation}
E \equiv E_n= -\hbar^2 \lambda\left(n + \frac{N}{2}\right)^2 + \hbar \left(n+\frac N 2\right ) \sqrt{\hbar^2 \lambda^2 \left(n+\frac{N}{2}\right)^2+ \om^2 } .
\label{re}
\end{equation}
Note that the degeneracy of this spectrum is exactly the same as in the $N$D isotropic oscillator, a feature that is again a signature of the maximal superintegrability of the model.

Concerning the continuous spectrum  of ${\hH}$, it can be rigorously  proven~\cite{BHERR}   that when $\la>0$, it is given by $[\frac{\om^2}{2\la },\infty)$. Moreover, there are no embedded eigenvalues and the singular spectrum is empty. The explicit form for the wave functions connected with the continuous spectrum can also  be given in hyperspherical coordinates:  it turns out that their  radial factor  can be expressed in terms of confluent hypergeometric functions \cite{BHERR}.

 On the other hand, it can also be   proven that ${\hH}$ has an infinite number of eigenvalues, all of which are contained in $(0,\frac{\om^2}{2\la })$ and their only accumulation point is $\frac{\om^2}{2\la }$, that is, the bottom of the continuous spectrum. 
 
 %%%%%%%%%%%%%%%%%%%%%%%%%%%%%%%%%%%%%%%%%%%%%%%%%%%

\subsection{Solution in hyperspherical  coordinates}

Let us now consider the Schr\"odinger equation (\ref{lh}) expressed in hyperspherical variables. This can be solved   
by factorizing the  wave function in the radial and angular components  and by 
 considering  the separability provided by the  first integrals   $\hat{C}_{(m)}$  with eigenvalue equations given by
$$
\Psi(r,\te)=\phi(r)Y(\te),\qquad \hat{C}_{(m)}\Psi = c_{m} \Psi,\quad m=2,\dots , N.
$$
From it, we obtain that $Y(\te)$ solves completely the angular part and is written, as expected, as the hyperspherical harmonics 
$$
\hbL^2 Y(\te)=\hbar^2 l(l+N-2)Y(\te),\quad l=0,1,2\dots  
$$
so that $l$ is the quantum number of the angular momentum.  The eigenvalues   of the operators  $\hat{C}_{(m)}$ are related with the  $N-1$ quantum numbers of the angular observables as 
$$
c_{k}\leftrightarrow l_{k-1},\quad k=2,\dots, N-1,\qquad c_N \leftrightarrow l ,
$$
 that is, 
$$
Y(\te)\equiv Y^{c_N}_{c_{N-1},..,c_2}(\theta_1,\theta_2,...,\theta_{N-1}) \equiv Y^{l}_{l_{N-2},..,l_1}(\theta_1,\theta_2,...,\theta_{N-1}) .
$$
Hence the radial Schr\"odinger equation reduces to
\begin{equation}
\frac 12 \left( {-\hbar^2}\left( \frac{\dd^2}{\dd r^2}  +\frac{N-1}{r} \frac{\dd}{\dd r}  -\frac{l(l+N-2)}{r^2}\right)+\omm^2 r^2\right)\phi(r) = E \phi(r) ,
\label{rad}
\end{equation}
where we have introduced the frequency (\ref{rc}). Since this is formally the same radial equation as for the $N$D isotropic oscillator,
we directly obtain the energy spectrum and the radial wave function:
\bea
&&E\equiv E_{k,l}=\hbar\omm\left(2k +l + \frac{N}{2}\right) ,\quad k=0,1,2\dots \nonumber\\
&& \phi(r)\equiv \phi_{k,l}(E,r) =B_{k,l} r^l \rme^{-\frac{\bb^2 r^2}{2 }}L_k^{(l+\frac{N-2}{2})}\left( {\bb^2 r^2}\right) ,\quad  \bb=\sqrt{\frac{\omm}{\hbar}}  ,
\label{rh}
\eea
where $L_{k}^{(l+\frac{N-2}{2})}$ are the associated generalized Laguerre polynomials and $B_{k,l}$ is a normalization constant. Finally, if we introduce the principal quantum number $n=2k+l$, that is, $E_{k,l}\equiv E_n$ (\ref{rree}) we recover the     energy spectrum (\ref{re}).

%%%%%%%%%%%%%%%%%%%%%%%%%%%%%%%%%%%%%%%%%%%%%

 \section{Generalization}
 
A quite simple but very important remark is worth to be commented. Let us consider any exactly solvable constant-mass Schr\"odinger problem with Hamiltonian
$$
\hat H = \frac{-\hbar^2}{2} \nabla^2 +\frac{\omega^2 }{2}\, U(\>q)
 \label{es}
$$
such that all the eigenvalues $E_n$ and eigenfunctions $\psi_n$ of the spectral problem
$
\hat H \,\psi_n=E_n\,\psi_n
$
can be obtained analytically.

Then, it turns out that the exact solvability of the position-dependent mass Hamiltonian $\hat H_\lambda$  defined by
$$
\hat H_\lambda = \frac{-\hbar^2}{2(1+\lambda\, U(\>q))} \nabla^2 +\frac{\omega^2 \, U(\>q)}{2(1+\lambda\, U(\>q))} ,
 \label{abs}
$$
is deeply related to that of $\hat H$, since the Schr\"odinger problem for $\hat H_\lambda$ reads
$$
\frac{-\hbar^2}{2(1+\lambda\, U(\>q))} \nabla^2 \psi_n^\lambda+\frac{\omega^2 \, U(\>q)}{2(1+\lambda\, U(\>q))}\psi_n^\lambda=E_n^\lambda\,\psi_n^\lambda ,
$$
and this equation can be written as
$$
\frac{-\hbar^2}{2} \nabla^2 \psi_n^\lambda+\frac{\Omega_\lambda^2 }{2}\, U(\>q)\,\psi_n^\lambda=E_n^\lambda\,\psi_n^\lambda ,
$$
which is just the spectral problem for $\hat H$ with the new energy-dependent frequency $\Omega_\lambda$ given by
$$
\Omega_\lambda=\sqrt{\omega^2-2\,\lambda\,E_n^\lambda}.
$$
Obviously, this spectral problem will present different features depending on the values of $\Omega_\lambda$, but in any case the exact solvability of $\hat H$ provides relevant information in order to get the eigenvalues and eigenfunctions for $\hat H_\lambda$. We stress that the addition of a nonzero constant to the potential $ U(\>q)$ (here scaled to 1) is essential in this procedure.

Therefore, the position-dependent mass system presented in this paper is the result of taking this $\lambda$-deformation approach when $U(\>q)$ is just the harmonic oscillator potential, thus explaining the maximal superintegrability of the system. Obviously, new $N$D radial models based on other well-known 1D  exactly solvable Hamiltonians can be constructed and will be presented elsewhere.

%%%%%%%%%%%%%%%%%%%%%%%%%%%%%%%%%%%%%%%%%%%%%

\section*{Acknowledgments}

This work was partially supported by the Spanish MICINN   under grants    MTM2010-18556   and FIS2008-00209, by the    Junta de Castilla y Le\'on  (project GR224), by the Banco Santander--UCM 
(grant GR58/08-910556) and by  the Italian--Spanish INFN--MICINN (project ACI2009-1083). 
   
   \newpage

%%%%%%%%%%%%%%%%%%%%%%%%%%%%%%%%%%%%%%%%%%%%%


\begin{thebibliography}{10}\frenchspacing

\small 
\bibitem{PhysD}
A.  Ballesteros,  A. Enciso, F.J. Herranz, O.  Ragnisco, {Physica D}   {237} (2008) 505.





\bibitem{Hietarinta}
J. Hietarinta, B. Grammaticos, B. Dorizzi, A. Ramani,  Phys. Rev. Lett. {53} (1984) 1707.


\bibitem{Stackel2}
E.G.  Kalnins,  J.M Kress,   W. Jr. Miller,
 {J. Math. Phys.} {46} (2005) 053510.

\bibitem{Stackel4}
E.G.  Kalnins,  J.M Kress,   W. Jr. Miller,
 {J. Math. Phys.} {47} (2006) 043514.




\bibitem{Sergyeyev}
A. Sergyeyev,     M. Blaszak,
J. Phys. A: Math. Theor. {41} (2008) 105205.


\bibitem{Kalnins1}
E.G.  Kalnins,     W. Jr. Miller, S. Post,
J. Phys. A: Math. Theor. {43} (2010) 035202.

 


\bibitem{IJTP}
A.  Ballesteros,  A. Enciso, F.J. Herranz, O.  Ragnisco,  D. Riglioni, Int. J. Theor. Phys.  (2010) {\em submitted},
arXiv:1010.3358.


 
\bibitem{annals}
A.  Ballesteros,  A. Enciso, F.J. Herranz, O.  Ragnisco,    {Ann. Phys.} {324} (2009) 1219.






\bibitem{Ko72}
G.  Koenigs,  {\em {L}e{\c{c}}ons sur la théorie générale des surfaces}
  vol.  4, ed.   G. Darboux, Chelsea, New York, 1972,  pp. 368.



  
\bibitem{KKMW03}
 E.G.  Kalnins,  J.M. Kress,  W. Jr.  Miller, P. Winternitz,
 {J. Math. Phys.} {44}  (2003) 5811.




\bibitem{PLB}
A.  Ballesteros,  A. Enciso, F.J. Herranz, O.  Ragnisco,   {Phys. Lett. B}  {652} (2007)   376.


\bibitem{Perlick}
V. Perlick,   {Class. Quantum Grav.}  {9} (1992) 1009.

\bibitem{Bertrand}
A.  Ballesteros,  A. Enciso, F.J. Herranz, O.  Ragnisco,    {Class. Quantum Grav.}  {25} (2008) 165005.

\bibitem{Bertrand2}
J.  Bertrand,   {C. R. Acad. Sci. Paris} {77} (1873) 849.




\bibitem{commun}
A.  Ballesteros,  A. Enciso, F.J. Herranz, O.  Ragnisco,      {Commun. Math. Phys.}  {290}  (2009) 1033.


 

\bibitem{IK95}
T.  Iwai, N.    Katayama,  {J. Math. Phys.} {36} (1995) 1790.




\bibitem{uwano}
T. Iwai,  Y. Uwano, N.  Katayama, {J. Math. Phys.} {37} (1996) 608.




\bibitem{Roos}
O.  von Roos,    {Phys. Rev. B} {27}  (1993) 7547.

 
\bibitem{mass2}
J.M.  L\'evy-Leblond,  {Phys. Rev. A} {52} (1995) 1845. 

\bibitem{mass3}
L. Chetouani, L. Dekar, T.F.  Hammann,  {Phys. Rev. A}  {52} (1995) 82.
 

\bibitem{mass4}
A.R. Plastino, A. Rigo, M. Casas, F. Gracias,  A. Plastino,  {Phys. Rev. A} {60} (1999)
4318.


\bibitem{Gritsev}
V.V. Gritsev,    Yu. A. Kurochkin, Phys. Rev.    B {64} (2001) 035308.


\bibitem{Quesnea}
C. Quesne, V.M. Tkachuk, {J. Phys. A: Math. Gen.} {37} (2004) 4267.


\bibitem{Quesne}
B. Bagchi, A. Banerjee,  C. Quesne, V.M. Tkachuk, {J. Phys. A: Math. Gen. } {38} (2005) 2929.

\bibitem{Koc} R. Koc, M. Koca, G. Sahinoglu, Eur. Phys. J. B, 48 (2005) 583.


\bibitem{Schd} A.G.M. Schmidt, Phys. Lett. A 353 (2006) 459.


\bibitem{Mustafa} O. Mustafa, S.H. Mazharimousavi, {Phys. Lett. A} {358} (2006) 259.

\bibitem{Quesnec}
C. Quesne,   {Ann.  Phys.} {321}  (2006) 1221.

\bibitem{mass5}
S.  Cruz y Cruz, J. Negro, L.M. Nieto,  {Phys. Lett. A} {369} (2007) 400.

\bibitem{SAG} A.G.M. Schmidt, A.D. Azeredo, A. Gusso, {Phys. Lett. A} {372} (2008) 2774.

\bibitem{MR}  B. Midya, B. Roy,
Phys. Lett. A 373 (2009) 4117.

 


\bibitem{BHERR}
A.  Ballesteros,  A. Enciso, F.J. Herranz, O.  Ragnisco,  D. Riglioni,    {Ann. Phys.} (2011)  {\em submitted}.

 
 
  
 
 
  
 



\end{thebibliography}
\end{document}